**Paper no. 18-06481**
**Utility Maximization Framework for Opportunistic Wireless Electric Vehicle Charging**

**MD Zadid Khan*, Ph.D. Student**
Glenn Department of Civil Engineering, Clemson University
351 Fluor Daniel, Clemson, SC 29634
Tel: (864) 359-7276, Fax: (864) 656-2670
Email: mdzadik@clemson.edu

**Mashrur Chowdhury, Ph.D., P.E., F. ASCE**
**Eugene Douglas Mays Endowed Professor of Transportation and**
**Professor of Civil Engineering and Professor of Automotive Engineering**
Glenn Department of Civil Engineering, Clemson University
216 Lowry Hall, Clemson, South Carolina 29634
Tel: (864) 656-3313, Fax: (864) 656-2670
Email: mac@clemson.edu

**Sakib Mahmud Khan, Ph.D. Candidate**
Glenn Department of Civil Engineering, Clemson University
351 Fluor Daniel, Clemson, SC 29634
Tel: (864) 569-1082, Fax: (864) 656-2670
Email: sakibk@g.clemson.edu

**Ilya Safro, Ph.D.**
School of Computing, Clemson University
228 McAdams Hall, Clemson, SC 29634
Tel: (864) 656-0637
Email: isafro@clemson.edu

**Hayato Ushijima-Mwesigwa, Ph.D. Student**
School of Computing, Clemson University
229 McAdams Hall, Clemson, SC 29634
Tel: (818) 428-9055
Email: hushiji@clemson.edu

**ACKNOWLEDGEMENTS**
This research is supported by the National Science Foundation under Award #1647361. Any opinions, findings, conclusions or recommendations expressed in this material are those of the authors and do not necessarily reflect the views of the National Science Foundation.

**INTRODUCTION**
Although Electric vehicles (EV) are more environment friendly in nature compared to conventional vehicles with internal combustion engines, they have some limitations. One of the limitations is range anxiety *(1)*, which can be solved by installing inductive coils in the bottom of EVs. EV batteries can be charged by inductive coils placed on the road when they are aligned with the inductive coils in the bottom of EVs. This is called dynamic charging or Charging While Driving (CWD). However, the efficient deployment of this technology largely depends on the effective infrastructure planning, where planners have to identify the strategic locations to place the Wireless Charging Unit (WCUs) on the road. Signalized intersections in urban areas can be an effective location to place WCUs as vehicles need to frequently stop at intersections.

In this paper, we present a framework titled "Simulation based Utility Maximization framework for Wireless Charging (SUM-WC)", which identifies the optimal placement of WCUs for a roadway network in an area of any size (for example, it could be a city, county or state). The objective of this research is to develop a method to identify the optimal locations to place the WCUs within a certain budget. The utility (i.e., total amount of energy in Watt-hour transferred to EVs from a WCU at a specific location within a certain timeframe) needs to be maximized while the corridor with WCU will continue to operate at an acceptable Level of Service (LOS). For signalized intersections, LOS signifies quality of service level based on the traffic control delay. Control delay can be described as the additional travel time encountered by a vehicle because of the presence of a traffic signal at an intersection *(2)*.

CWD for EVs has been studied with a focus on design and implementation of efficient charging systems *(3, 4)*. The authors in *(5)* have designed and implemented an optimized system for wireless charging for an electric bus that travels on a fixed route. Other studies *(6-8)* have investigated the location of deployment for WCUs. For example, in *(8)*, authors have introduced an integer-programming model that simulates different realistic scenarios of EV routes. The authors have compared the computational results for the proposed model with faster heuristics and demonstrated that their approach provides better results for fixed budget models. However, these studies do not consider the effect of traffic signals on EV charging in the analysis. The authors in *(9)* have investigated improvement of fast charging process for connected EVs, and the authors in *(10)* have investigated dynamic routing of connected EVs. The authors in *(11)* have studied wireless charging at signalized intersections using simulation; the objective of this study is to meet EV's energy demand and minimize delay at intersections. However, the analysis comes more from the traffic signal perspective; adaptive control strategies of traffic signals are developed for charging of EVs at intersections. Based on the reviewed literature, it can be stated that the study in this paper is unique as it incorporates traffic signal control strategies to determine the optimal location of WCUs.

**METHOD**
The framework for our study is based on a single objective optimization formulation, where the decision variable are, (a) the placement and sizing of WCUs, and (b) the traffic signal timings. The objective function is the total utility of the deployment, and it is a maximization problem. The constraints on the decision variables are based on control delay and budget. To the best of our knowledge, this is the first study that models the interplay between opportunistic wireless charging at signalized intersections and the traffic signal operation, such that the utility of WCU deployment is maximized within a given budget.

**Assumptions**
All signals are assumed to be two phase pre-timed signals, and the cycle time is assumed to be fixed. So, an increase in green time of one phase will result in the decrease of green time in the other. While fixed-time signals are not required for the framework, it simplifies the optimization problem formulation and analysis. The framework can be extended to include other signal control strategies (e.g., actuated, semi-actuated or adaptive).

**Modeling Approach**
The first step is to build a traffic simulation model for the area of interest. The model is then simulated once with no WCU to obtain an average control delay for all lane groups at all intersections (i.e., a lane group consists of lanes that share a common stop bar at intersections). The set of lanes approaching the traffic signals is then identified. The lanes with LOS D, E and F are omitted from the analysis, since they are performing at an unacceptable LOS *(2)*. Moreover, the lanes with LOS A and B are characterized by a low control delay that is not sufficient to charge the EVs properly while stopped. We thus select lanes from the lane group that have LOS C. We next simulate the pre-built traffic simulation model with WCU placed at each of the selected lanes. We simulate multiple scenarios with low-mid-high values (minimum allowable value, base value and maximum allowable value) of the number of WCU units at each lane and low-mid-high time for each signal phase of each traffic signal. The total energy transferred to all EVs (i.e., the utility) and average control delay of all selected lanes for each simulation is subsequently extracted. An interpolant is developed for utility of each selected lane depending on the number of WCU units and time of each phase. Similarly, an interpolant is developed for control delay of each selected lane depending on the time of each phase. The optimization problem is solved using Genetic Algorithm, which was selected because it can solve both single and multi-objective mixed integer non-linear optimization problems *(12)*. The effectiveness of the optimization framework is then compared with the betweenness centrality scheme, which creates a ranking of the edges based upon their significance *(13)*. The significance of an edge can be quantified as the number of shortest paths that include this edge, this value is known as the "betweenness centrality" of that edge.

**FINDINGS**
Here, the evaluation of the SUM-WC framework for a sample road network is described, with the network, its properties and the solution detailed in FIGURE 1.
    The constraints for this analysis must first be specified, with the budget set at $10,000, and cost of one unit set at $2,000. Therefore, five units are available for installation. The minimum green time for any phase is set at 15s. LOS of all intersections must be either better than or equal to LOS C.

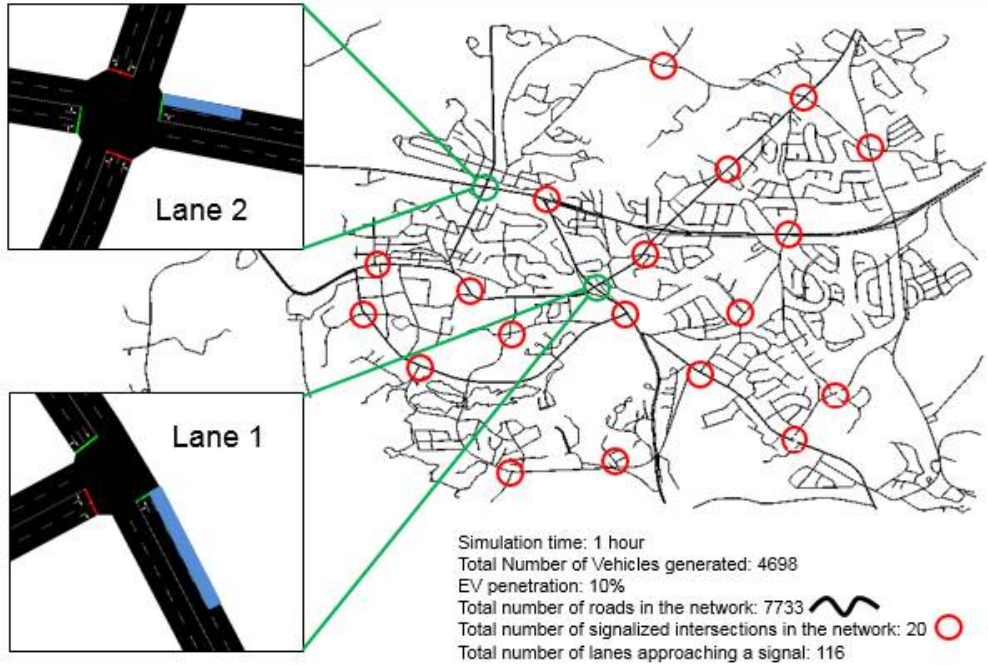

FIGURE 1. Sample network with the solution

The simulated network has 20 intersections and 116 lanes in which WCUs can be installed. Since 42 of these lanes have LOS C, the utility and control delay interpolants are formulated for the 42 lanes according to the SUM-WC framework. The optimization is then solved to obtain the global solution with the Genetic Algorithm. Two lanes (Lane 1 and Lane 2) from two different intersections (Int. 1 and Int. 2) are selected for WCU installation with three units installed at Lane 1 and two units at Lane 2. The green times of the two phases at Int. 1 are 22s and 35s, and the green times of the two phases at Int. 2 are 17s and 40s. Both intersections operate at LOS C. The total utility achieved is 3KWh, which is the total charging of the group of EVs that pass over the WCUs installed in both lanes. These results are compared with the betweenness centrality scheme for allocating the WCUs in the road network. The two most significant edges (edges with the highest betweenness centrality values) are chosen from the network. Three units are installed in one lane, and two units are installed in the other, no changes are made to signal times from the base case. The total utility achieved through this scheme is 2 KWh, compared to 3 KWh from SUM-WC framework. The SUM-WC produces 1.5 times the utility rate of betweenness centrality.

**CONCLUSIONS**
The SUM-WC framework is successful in achieving a 1.5 times greater utility rate compared to the betweenness centrality scheme, indicating its use as the foundation for future research on opportunistic CWD analysis.

There are a few limitations of this analysis, which can be addressed in future research. Actuated or adaptive signal controls is not considered, the control delay is not optimized at a network level, and no other deployment schemes have been considered for comparison with the SUM-WC framework except the betweenness centrality scheme. These issues can be addressed in future analysis. After successful installation of WCUs in the network, the real time data sharing will become very important. Connected EVs can communicate with the infrastructure for smart decision-making. Therefore, elucidating the efficacy of connected vehicle technology in providing EV users with efficient charging strategies could also be the subject of future research.